# Influence of Disorder on Exciton Transfer in a Quantum Dot Chain with Short-Range Interaction and a Side-Coupled Defect

A. D. Vlasov, P. A. Golovinski

Voronezh State University

University Square, 1, Voronezh 394018, Russian Federation

**Abstract**

Colloidal quantum dots (QDs) can be efficiently synthesized in solution, and when deposited on surfaces, they form aggregates of various types. This paper considers the propagation of excitons in linear chains of QDs with a side defect, located on a dielectric substrate. This configuration is suitable for spatially selective excitation of the system by pulsed optical radiation through the side defect. The dynamics of excitation in the chain is governed by structural disorder, caused by technological variations in the parameters of the QDs themselves and their mutual arrangement. To describe the quantum properties of excitons in the QD chain, a model Hamiltonian is used, taking into account the coupling of neighboring QDs due to dipole–dipole interaction. The localization of stationary states is calculated depending on the magnitude of disorder and the chain length. A criterion is introduced that determines the boundaries of the phase transition from the localized to the delocalized excitation phase. The dynamics of exciton transfer along the QD chain is calculated depending on its length and degree of disorder for linear excitation of the system by a laser pulse. Dissipation is incorporated via a non-Hermitian term in the Hamiltonian. It is shown that dynamic localization emerging in the system corresponds to the stationary localization properties of the states of the chain with a side defect.

**Keywords:** colloidal quantum dots, quantum dot chain, side defect, disorder, Anderson localization, ultrashort pulse, excitation propagation, quantum transport.

1. **Introduction**

Quantum dots (QDs), or "artificial atoms," are nanometer-sized semiconductor crystals [1-3] with three-dimensional confinement of electron motion, leading to a discrete energy spectrum and allowing control over their optical and electronic properties. Arrays of QDs are a promising platform for quantum information processing, enabling manipulation of ensembles of coherent quantum states [4-7] using metallic Stark gates or laser radiation.

Excitons in QDs are becoming a sought-after physical basis for quantum computing circuits. The advantages of such circuits are related to the possibility of laser excitation of excitons and their coherent transport between QDs through resonant Förster dipole–dipole interaction without charge transfer, which eliminates ohmic losses. Resonant exciton transport has been observed experimentally [8] and proved significantly more efficient in superlattices than in disordered films, emphasizing the role of disorder in reducing coherence [9].Of particular interest are one-dimensional QD chains with a side-coupled structure, possessing unique transport and spectral properties. The possibility of forming such chains on dielectric substrates [10, 11] makes them technologically accessible. Local generation of excitons in individual QDs can be provided in the near field by a short pulse [12] through surface plasmon-polariton resonance on a waveguide in the form of a nanoneedle or thin silicon film [13], allowing the optical field to be focused below the diffraction limit.

Stationary and dynamic properties of excitons in ordered QD chains of various topologies are actively studied within the single-particle strong-coupling model [14], including the influence of monochromatic, pulsed laser fields or the Stark effect. Significant technological variation in QD sizes and distances between them is inevitable in colloidal synthesis and surface deposition, leading to random variations in energy levels and Förster interaction. Localization of quantum states in disordered systems was first described by Anderson [15], and statistical properties of finite one-dimensional chains without side structures have been investigated for diagonal [16] and off-diagonal disorder [17].

In this work, a one-dimensional disordered QD chain of finite length with a side defect in the form of a single QD is investigated. The stationary energy spectrum and dynamics of system excitation by a short laser pulse through the defect are numerically modeled and analyzed, depending on the statistical characteristics of the ensemble of such chains.

## 2. Mathematical Model

To construct the model, the Anderson approach is used, based on a single-particle description of states and the strong-coupling approximation. The quantum state is described by probability amplitudes at the nodes of the QD chain. Only the lowest single-exciton states with energies $E_n$ are considered. A high potential barrier between QDs is assumed, excluding electron tunneling but allowing an effective Förster interaction. Structural variations are modeled by random diagonal (site energies $E_n$) and off-diagonal (coupling constants $V_{n,n+1}, V_{n+1,n}$) elements of the Hamiltonian. To describe excitation by an external optical pulse $\mathcal{E}(t)$, a side QD is introduced, considered as a two-level system with ground-state energy $E_0$ and excited-state energy $E_d$. Resonant excitation of the side defect is transferred to the chain through its coupling with the $k$-th point of the chain, which is the nearest to the defect. The excitation transfer scheme in the system is shown in Fig. 1.

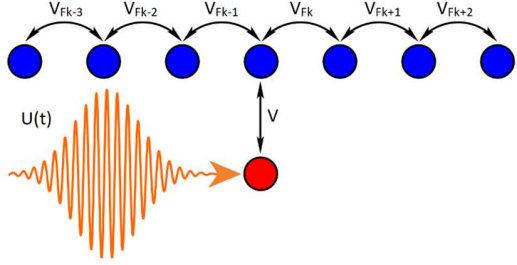

*Fig.1. Scheme of pulsed laser excitation and exciton transfer in a QD chain with a side defect*

According to the chosen model, the non-stationary nodal Hamiltonian of the QD chain with a side defect, written in matrix form, is
$$H(t) = H_0 + i\Gamma + V(t), \qquad (1)$$
where $H_0$ is the Hermitian part, $\Gamma$ is the dissipation matrix (diagonal with elements $\gamma_n$). Constants $\gamma_n$ (approximately the same for all QDs) set the level widths and describe relaxation of states due to exciton interaction with the environment. The perturbation operator $V(t)$ in Hamiltonian (1) contains the product of the electric field strength $\mathcal{E}(t)$ and the matrix element of the dipole moment operator $d$: $V(t) = d \cdot \mathcal{E}(t)$. In the weak-field (linear response) regime, the system response is proportional to the external field amplitude, so the dynamics can be studied independently of the field amplitude by normalizing. At $V(t) = 0$, Hamiltonian (1) allows determining the stationary states of the system and their energies as the solution to the system of equations
$$H_0|\psi\rangle = E|\psi\rangle. \qquad (2)$$
We use the basis $\{\psi_n\}, \psi_d$, where $\psi_n$ is the wave-function amplitude of the chain in nodal representation corresponding to the $n$-th QD (node), $\psi_d$ is the excitation amplitude of the side QD.

Solving this system of equations yields eigenvalues $E_j$, corresponding state vectors $|\psi^j\rangle$, and probability distributions $P_n^j = |\psi_n^j|^2$ of finding the particle at nodes of the system corresponding to the $j$-th eigenenergy. In the absence of disorder for an infinitely extended system, the eigenfunctions of equation (2) without side defect are Bloch waves. With the emergence of disorder in an infinite one-dimensional chain, the eigenfunctions are guaranteed to localize in random regions of the system, and the energy spectrum becomes disordered. Localization of states is characterized by a typical length $L$, determining the size of the system region where the probability density of the eigenstate is concentrated, and it decreases with increasing disorder in the system.

### 3. Results of Numerical Calculations and Discussion

#### 3.1. Statistics of Localized States

Despite a large number of studies devoted to Anderson localization, there is currently no complete mathematical description of this phenomenon for finite-length chains. Therefore, in this work, numerical modeling is used as a research tool. We begin with a statistical analysis of the distributions of localization lengths $P(L)$. Various methods exist for calculating the localization length $L$, developed for analysis in the limit $N \to \infty$, such as computing the Lyapunov exponent [18] or the IPR (inverse participation ratio) indicator [19]. In our work, a method for calculating $L$ convenient for finite $N$ values is applied, where localization in the main chain nodes is controlled by components of eigenstates. The localization length is defined as the spatial extent of the wave function between the outermost sites with non-negligible probability density. This definition is numerically stable for finite chains and allows direct characterization of spatial support of eigenstates without assuming exponential decay. The localization length $L_j$ of a given eigenvector is the difference of the corresponding indices. Statistical numerical modeling of an ensemble of random stationary Hamiltonians for system (1) was conducted with given standard deviations $\sigma_E$ and $\sigma_V$ for matrix elements of the main and secondary diagonals, used as a measure of disorder in the initial Hamiltonian. For each matrix in the ensemble, diagonalization of equation (2) was performed, resulting in eigenvectors necessary for constructing the distribution of localization lengths. These distributions allow evaluating the degree of localization of eigenstates for the ensemble of realizations at given disorder of main diagonal elements $\sigma_E$ and secondary diagonals $\sigma_V$. The case of identical simultaneous disorder of the main diagonal and secondary diagonals with Gaussian distribution of elements was considered, i.e., $\sigma = \sigma_E = \sigma_V$.

For numerical modeling, initial parameters corresponding to CdSe QDs of about 2 nm size with ground excitonic state energy of about 2 eV were used [2]. The binding energy due to resonant Förster interaction of two neighboring QDs depends on the distance between them and is approximately 0.3 meV [20] at a distance of 2 nm. These characteristics were taken as average values of the diagonal $E$ and off-diagonal $V$ elements of Hamiltonian (1). The average interaction energy $V_d$ of the defect with the $k$-th chain point was taken equal to $V$. The number of realizations of stationary Hamiltonians in the ensemble is chosen to ensure a stable form of the calculated probability distributions $P(L)$. In the considered modeling example, the ensemble size was $10^4$ matrices. The chain length was set to $N = 301$ QDs, and the side defect is located in the center, at $k = 151$. The results of numerical modeling show that at small disorder values, the predominant extension of eigenvectors is comparable to the size of the entire system (Fig. 2a), i.e., localization is expectedly not observed. With increasing disorder, localization lengths become almost equally probable, indicating a transition to a stage of chaotic distribution of localized states (Fig. 2b) with manifestation of localization on the defect. With further increase in disorder, a transition to localization lengths (Fig. 2c) smaller than the chain length occurs. Thus, the order–disorder transition is conveniently characterized by two integral characteristics: the average value of the localized state length $\langle L \rangle$ and the standard deviation of lengths $\sigma_L$.

The parameter region where $\langle L \rangle \approx N$ and $\sigma_L \ll N$ corresponds to the delocalization phase covering the entire chain. The region of almost uniform distribution of state lengths, where $\langle L \rangle \approx N/2$, is the transition phase, and the region of small $\langle L \rangle$ with $\sigma_L \ll \langle L \rangle$ is the localization phase. Thus, the change in distribution can be tracked by $\sigma_L$. With increasing disorder, $\sigma_L$ values first grow, corresponding to increased fluctuations, and then decrease upon transition to the localized phase.

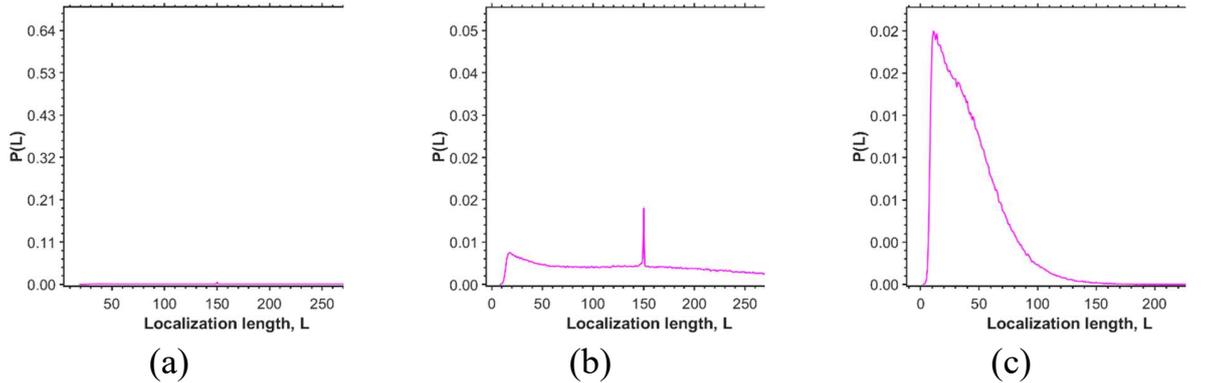

(a)            (b)            (c)

*Fig2. Distributions $P(L)$ for various disorder magnitudes of diagonals in the case $\sigma_E = \sigma_V = \sigma$: (a) – $\sigma = 10^{-6}$ eV; (b) – $\sigma = 10^{-4}$ eV; (c) – $\sigma = 10^{-2}$ eV*

Note that in the limit of an ordered chain ($\sigma = 0$), the coupling of the side defect to the chain affects the spectrum depending on the matrix element magnitude $V_d$. At weak coupling, the side QD practically does not affect the form of eigenvectors. At $V_d \approx V$, the vectors split into three types: with $L = k$, $L = N - k$, and $L = N$. Further growth of $V_d$ leads to an increase in the probability of finding states with $L = k$, $L = N - k$ among the eigenvectors. Thus, at strong coupling of the side QD to the chain, two types of eigenvectors appear in the spectrum: some have $k$ components close to zero and $N - k$ delocalized components, others – vice versa. Thus, a localization effect appears, not related to disorder, but arising due to hybridization of the side-node states and the chain.

Analysis of the behavior of quantities $\langle L \rangle$ and $\sigma_L$ when varying disorder parameters allows establishing their general dependence on dispersions of diagonal energies and Förster interaction. This provides a clearer understanding of the phase transition boundaries, as shown in Fig.3a,b. The ensemble size used for calculating $\langle L \rangle$, $\sigma_L$ at each point $(\sigma_E, \sigma_V)$ is 1000 matrices; the step in disorder magnitude of diagonals $\Delta\sigma = 10^{-5}$ eV.

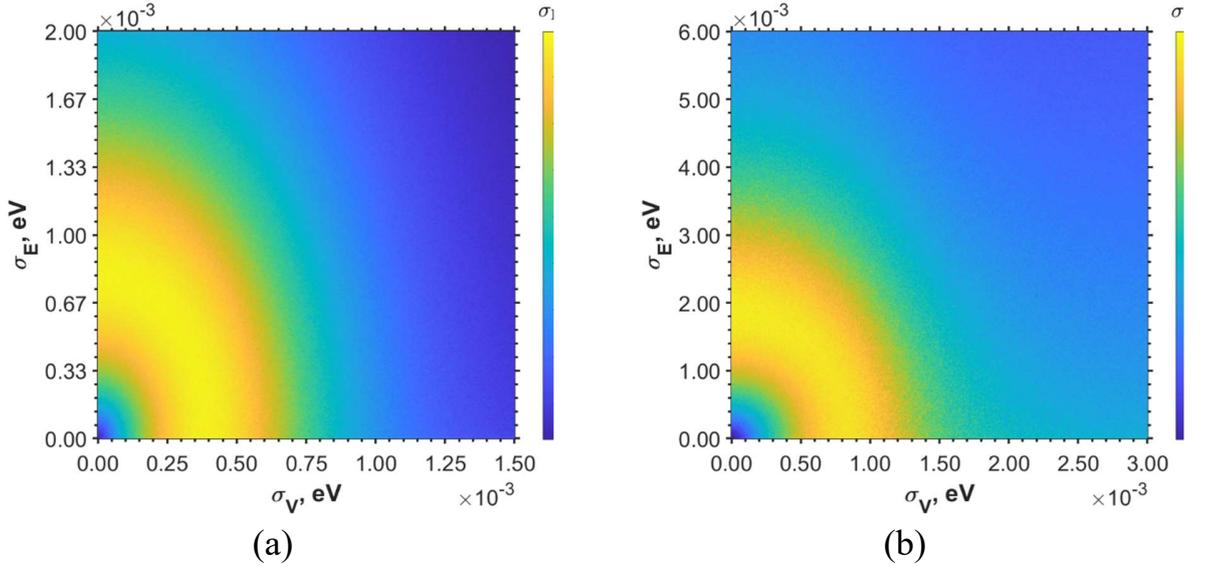

(a)          (b)

Fig.3. Distributions $\langle L \rangle$, $\sigma_L$ for chains with side defects located in the center: (a) – $N = 301$; (b) – $N = 81$

The obtained three-dimensional surfaces are presented as color maps of $\langle L \rangle$ and $\sigma_L$ dependences on $(\sigma_E, \sigma_V)$. The color scales indicate the values of $\langle L \rangle$ and $\sigma_L$. The inner boundary (small blue ellipsoid) separates the region of initial delocalization of states, where fluctuations in localization lengths are small, from the transition phase (yellow region), in which localization lengths are distributed almost uniformly across nodes and have large standard deviation. The outer ellipsoidal boundary separates this transition phase from the strong localization region (blue color), where the standard deviation is again minimal.

The calculated $\langle L \rangle$ and $\sigma_L$ show the size effect: for shorter chains at fixed disorder magnitude, a shift in phase transition boundaries is observed. At a disorder value sufficient for transition of a long chain (Fig. 3a) to the localized phase, a short chain still remains in the delocalized phase (Fig. 3b). For sufficiently long chains, when $N \gg L$, $L$ is determined only by the degree of diagonal disorder. At fixed disorder values, the value of $L$ remains constant regardless of further growth of $N$.

The boundaries of disorder parameters for the existence of different phases are conveniently approximated by elliptical curves. The condition for localization of excitations takes the form

$$\left(\frac{\sigma_E}{a}\right)^2 + \left(\frac{\sigma_V}{b}\right)^2 > 1, \tag{3}$$

where $a$ and $b$ are parameters linearly dependent on the average value of the dipole–dipole coupling magnitude $V$. The possibility of excitation localization in the QD chain can be established by calculating $a$ and $b$ for a specific system. Having such diagrams, one can see the boundaries of "order – disorder" phases depending on the system disorder parameters.

### 3.2. Dynamics of Chain Excitation by Laser Pulse Field

To reveal localization features in the dynamics, we consider exciton transport caused by excitation of the side defect by an ultrashort laser pulse. We introduce the transition matrix element in the two-level system of the side defect in the form of a pulsed time dependence with Gaussian envelope:

$$d(t) = d_0 \exp\left(-\frac{(t-t_0)^2}{2\tau^2}\right)\cos(\omega t), \tag{4}$$

Here $d_0$ is the amplitude of the excitation matrix element of the side defect, $\tau$ is its duration, $\omega$ is the carrier frequency. Parameter $\tau = 31$ fs ensures $\sim 15$ oscillations; $\omega = 2\pi f$, $f$ taken resonant and equal to 483.6 THz, corresponding to energy 2 eV. Exciton state lifetime $T = 1$ ns, and its width $\gamma = \hbar/T \approx 0.66$ $\mu$eV. Dynamics calculations were performed at $d_0 = 0.05$ eV, ensuring linearity of the field impact on the system. Initial conditions were chosen as $\psi_0(0) = 1$, $\psi_d(0) = 0$, $\psi_n(0) = 0$ and correspond to the unexcited state of the side defect and the entire linear chain. Numerical solution of the non-stationary Schrödinger equation with Hamiltonian (1) was carried out using the Crank–Nicolson method [21], in which the change in system state per integration step is determined by the relation

$$|\psi(t + \Delta t)\rangle = (I + iH\Delta t/2)^{-1}(I - iH\Delta t/2)|\psi(t)\rangle, \qquad (5)$$

where $I$ is the identity matrix, $\Delta t$ is the time step. This method is stable for calculating quantum dynamics. For modeling, statistical system parameters used in section 3.1 were used. In linear mode, the probability $P_n(t)$ (node population) is proportional to the square of the external field amplitude, i.e., $P_n(t) \propto |d_0|^2$. At the same time, the population of the initial ground state of the side defect remains close to unity ($P_0 \approx 1$), and populations of other states remain small $P_d, P_n \ll 1$. Thus, the chosen linear mode is realized.

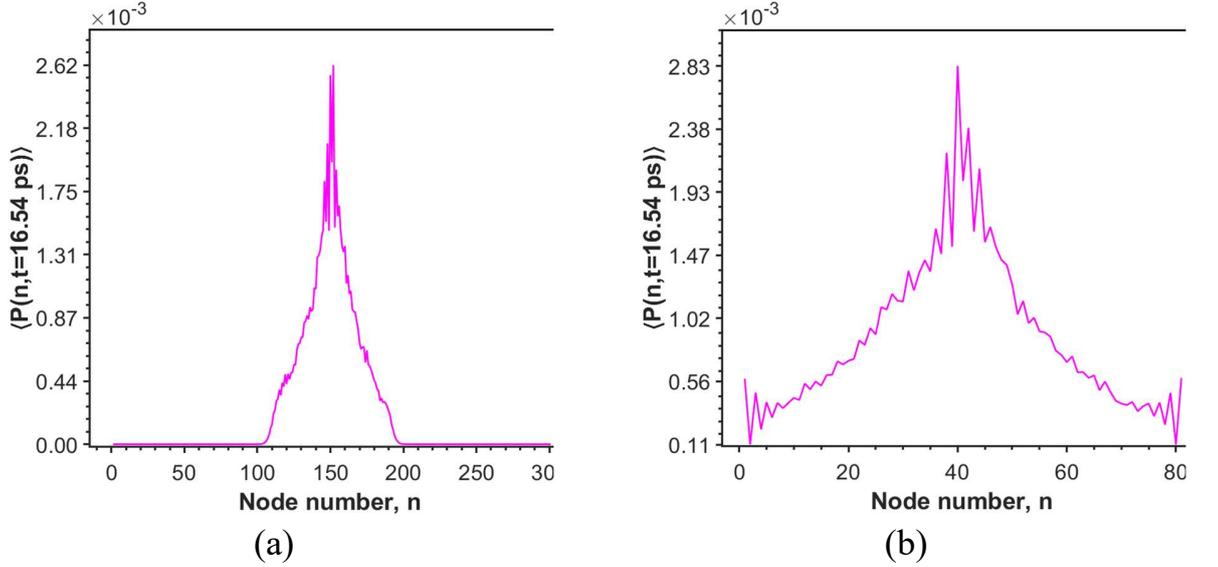

(a)            (b)

*Fig.4. Average probability density of node wave-function excitation for chains with side defects located in the center: (a) – $N = 301$; (b) – $N = 81$*

Numerical modeling of the dependence of dynamic localization of the wave packet on chain length $N$ showed that for the same disorder magnitude for a short QD chain, excitation reaches the edges, while for a longer chain, it localizes in some region. As seen from Fig. 4a, at disorder magnitude $\sigma = 10^{-4}$ eV, identical for diagonal and off-diagonal elements, the maximum chain length at which excitation still reaches the edge is ~80. In such a chain, excitation is in the transition phase from order to disorder. At larger $N$, as shown in Fig. 4b, the wave packet does not reach the boundaries, which are reached in the case of a shorter chain. The established dynamic properties are actually a consequence of the previously described characteristics of stationary localization in finite-length chains. Disorder controls the dynamics of excitonic transport, leading to dynamic localization of the wave packet and the appearance of a trapped, localized wave.

## 4. Conclusion

We have shown that increasing disorder in a linear QD chain leads to localization of exciton states, demonstrating signatures of the Anderson transition in a finite one-dimensional QD chain. An analytical form of the ellipsoidal boundary criterion for the phase transition from delocalized exciton states to localized ones is established.

The dynamics of chain excitation through the side defect under the action of an ultrashort resonant laser pulse is analyzed. Numerical modeling revealed correspondence of the chain excitation dynamics to the phase state of its stationary states: at weak disorder, excitation reaches the chain edges, while at wave function localization – it is trapped inside, suppressing coherent transport across the chain. Transport efficiency is shown to be governed by the defect–chain coupling strength: at strong coupling, hybridization with nearest QDs occurs and localization near the defect, not related to disorder. Numerical modeling was conducted over a wide range of chain length parameters, and it confirms the results presented in the article for specific examples.

Exciton transport can be experimentally investigated, for example, by time-resolved photoluminescence method [22]. As a potential application, we note that linear chain with a side defect, excited through a nanoneedle or waveguide, can serve as a quantum modulator. Controlled shift of QD levels, for example, by applying a set of different potentials to nanoelectrodes near QDs, changes the spectral response of the chain to laser pulse action, which is equivalent to modulation of wave function amplitudes in energy representation. The technical feasibility of realizing such structures is demonstrated in [23].

Further theoretical studies of QD chain dynamics for creating quantum modulators should include calculation of transmission spectra accounting for both disorder in the chain with long-range interaction and signal modulation depth, which will allow quantitative assessment of the bandwidth of such systems.

### Authors' Declared Contribution

All authors contributed equally to this work.

### Conflict of Interests

The authors declare that they have no known financial conflicts of interest or personal relationships that could have influenced the work presented in this article.